# Uma técnica para a quantificação do esforço de merge

**Tayane Silva Fernandes de Moura, Leonardo Gresta Paulino Murta**

Universidade Federa Fluminense (UFF) – Instituto de Computação
Av. Gal. Milton Tavares de Souza, s/nº São Domingos - Niterói – RJ CEP: 24210-346

`tayanemoura@id.uff.br, leomurta@ic.uff.br`

***Abstract.*** *Developers that use version control systems can work in parallel with other developers and merge their versions afterwards. Sometimes these merges fail, demanding manual intervention to resolve conflicts. Some studies aim at analyzing the merges failures, however, there is a lack of tool support in the literature to measure the merge effort, jeopardizing such kind of analyses. In this article, we propose a technique and its companion tool for analyzing Git repositories and providing metrics related to the merge effort. We evaluated our tool over five projects, showing that rework and wasted work happens in, approximately, 10% to 30% of the projects. Moreover, the average of actions of these efforts is almost the same.*

***Resumo.*** *Com o uso de sistemas de controle de versão os desenvolvedores podem trabalhar em paralelo e, ao final, integrar (merge) suas contribuições. Algumas vezes esses merges falham e é necessário a intervenção de um desenvolvedor para resolver os conflitos. Existem estudos que buscam analisar essas falhas de merge e, para isso, necessitam quantificar o esforço envolvido. Esse artigo propõe uma técnica, implementada em uma ferramenta, capaz de analisar repositórios Git e fornecer métricas relacionadas ao esforço de merge. A ferramenta proposta foi avaliada em cinco projetos e pudemos observar que a incidência de retrabalho e de trabalho desperdiçado acontece em, aproximadamente, 10% a 30% dos commits de merge. Além disso, a quantidade média de ações desses esforços é muito parecida.*

## 1. Introdução

Sistemas de Controle de Versão (SCV) são ferramentas que controlam mudanças nos artefatos de software, denominados Itens de Configuração (IC) [Grinter 1995]. Além disso, os SCV permitem acesso aos repositórios em que os artefatos e seus históricos estão guardados e também proporcionam mecanismos para bloquear, comparar e combinar (*merge*) diferentes versões do mesmo IC, o que possibilita o desenvolvimento paralelo de um software. Quando é utilizada a política otimista de controle de concorrência, dois artefatos podem ser alterados simultaneamente, o que muitas vezes pode agilizar o desenvolvimento. Ao final das alterações, as duas versões são combinadas (i.e., *merge*), gerando-se uma nova versão do IC [Prudêncio et al. 2012]. Estudos passados indicam que 10% a 20% dos *merges* falham [Brun et al. 2011; Kasi and Sarma 2013] e quando isso ocorre os desenvolvedores precisam fazer manualmente o *merge* dessas versões.

Existem estudos que buscam analisar os conflitos de *merge*. Accioly et al. [2017] analisaram os conflitos de *merge* e elaboraram, através do estudo de mudanças

de código, um catálogo de padrões de conflitos. Santos et al. [2012] também fizeram um estudo em torno de características de *merges*. Nesse estudo, os autores apresentam uma abordagem para a extração de métricas que visam prever se o *merge* de ramos será difícil. Existem ainda outros estudos que visam entender a natureza dos conflitos de *merge* [Kasi and Sarma 2013; Menezes 2016; Yuzuki et al. 2015]. Todos esses estudos se beneficiariam de uma técnica capaz de quantificar o esforço envolvido nos merges.

Alguns trabalhos [Cavalcanti et al. 2017; Mehdi et al. 2014; Prudêncio et al. 2012; Santos and Murta 2012] fazem uso de métricas de esforço de *merge* nas suas análises, porém, como o foco desses trabalhos não é na extração de métricas de esforço, a descrição da técnica de extração adotada é usualmente superficial e a implementação dessa técnica raramente está disponível de forma desacoplada para uso por outros pesquisadores. Assim, a cada pesquisa no tema, os pesquisadores se veem obrigados a reimplementar seus próprios coletores de métricas de *merge*. Como efeito colateral, há um elevado grau de retrabalho e pouca certeza que os resultados dos estudos possam ser comparados, já que cada implementação atua numa granularidade distinta e segue uma técnica distinta. Como consequência da pouca transparência sobre o funcionamento de cada uma dessas técnicas, a interpretação e a reprodução dos resultados são prejudicadas.

Nesse artigo, propomos uma técnica, e a sua implementação em uma ferramenta, para apoiar pesquisadores nos estudos sobre o esforço de *merge*. A partir da URL de um repositório Git, a ferramenta é capaz de identificar todos os *commits* de *merge* do repositório e calcular o trabalho extra necessário para efetuar cada *merge* (código que pertence ao *commit* de *merge* mas não veio dos ramos), assim como o retrabalho (código duplicado nos ramos) e o trabalho desperdiçado (código que pertence aos ramos e que não foi incorporado no *merge*). Cada uma dessas métricas é calculada de forma absoluta e normalizada, atendendo assim às várias necessidades de uso.

Para avaliar a ferramenta fizemos uma análise em cinco projetos open-source e concluímos que a incidência de retrabalho e de trabalho desperdiçado acontece em aproximadamente 10% a 30% dos *commits* de *merge*. Além disso, a quantidade média de ações desses esforços é muito parecida.

O restante do artigo está organizado em outras cinco seções. Na Seção 2, detalhamos a técnica proposta, explicando o processo de merge e a extração das métricas de esforço. Na Seção 3, exploramos a implementação da nossa ferramenta. Na Seção 4, apresentamos os resultados obtidos na avaliação da ferramenta. Na Seção 5, apontamos os trabalhos relacionados. Por fim, na Seção 6, expomos nossa conclusão e trabalhos futuros.

## 2. Esforço de Merge

Em um SCV, dois desenvolvedores podem trabalhar em paralelo no mesmo IC e depois combinar suas versões. Para ilustrar esse processo, criamos o cenário da Figura 1. Inicialmente, o desenvolvedor cria o ramo *dev* a partir do *commit base* C3 e realiza algumas modificações no código, gerando os *commits* C4, C6 e C7. Enquanto isso, um outro desenvolvedor continua trabalhando no ramo *master*, produzindo os *commits* C5 e C8. Ao final das modificações é feito o *merge* entre os ramos, gerando o *commit de merge* C9, a partir dos *commits pais* C7 e C8.

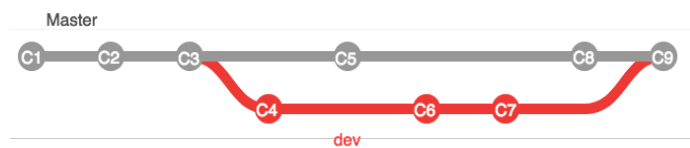

**Figura 1: Exemplo de merge entre ramos**

Quando o *merge* falha, o desenvolvedor precisa resolver o conflito manualmente. Quando o *merge* manual é necessário, o desenvolvedor deve escolher entre uma das duas versões, concatenar as duas versões em alguma ordem, combinar as duas versões de alguma forma ou escrever um novo código [Menezes 2016].

Um estudo [Menezes 2016] mostrou que 75% dos conflitos ocorridos nos repositórios estudados foram resolvidos escolhendo-se uma das duas versões. Quando o desenvolvedor opta por uma versão em detrimento da outra, existe um *trabalho desperdiçado* [Prudêncio et al. 2012], já que houve a criação de código em um dos ramos que não foi incorporado no *merge*. Na Figura 2 podemos ver que o trabalho desperdiçado é o código de uma ação que foi realizada, mas que não foi integrada durante o *merge*. No mesmo estudo, Menezes [2016] concluiu que 13% dos conflitos foram solucionados com código novo, ou seja, demandando um esforço adicional durante o *merge*. Denominamos esse esforço adicional como *trabalho extra*, que consiste nas ações que foram integradas ao código, mas que não foram realizadas anteriormente ao merge. Além disso, quantificamos também o *retrabalho*, que acontece quando existe interseção entre as ações criadas em paralelo, como na Figura 2.

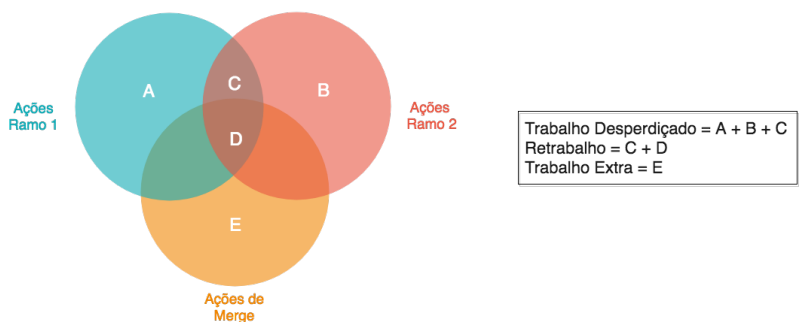

**Figura 2: Tipos de métricas**

Para representar cada conjunto das ações, utilizamos *multisets* [Knuth 1998], tendo em vista que pode haver repetições de ações nos conjuntos. Cada *multiset* contém as ações que representam as adições ou remoções de linhas de código. Os *multisets* são compatíveis com as operações usuais de conjuntos, como diferença, interseção e união, além de adicionar uma operação de soma. Por exemplo, supondo $A = \{a, a, b, c, c\}$ e $B = \{a, b, b, c, c\}$, temos $A \backslash B = \{a\}$, $B \backslash A = \{b\}$, $A \cap B = \{a, b, c, c\}$, $A \cup B = \{a, a, b, b, c, c\}$ e $A + B = \{a, a, a, b, b, b, c, c, c, c\}$.

Com o intuito de extrair as métricas, é necessário primeiro fazer a análise dos *commits* do repositório a procura dos *commits* de *merge*. Para isso, verificamos cada um dos *commits* em busca daqueles que têm dois pais. Como merges com mais de dois pais, denominados *octopus*, não podem ter conflitos e edições manuais por definição, esses casos teriam zero esforço e são ignorados. A partir de um *commit* de *merge* denominado $commit_{merge}$, são obtidos os seus *commits* pais e, a partir deles, é identificado o

*commit* em que eles derivaram, denominado $commit_{base}$. Diante dessas informações, são realizadas três operações de *diff* para obter as referidas ações. O primeiro *diff* é realizado entre $commit_{base}$ e $commit_{merge}$, obtendo-se assim as ações que foram de fato incorporadas no *merge*. Definimos formalmente as ações de *merge* como:

$$actions_{merge} = diff(commit_{base}, commit_{merge})$$

Em seguida, é realizado o *diff* entre $commit_{base}$ e $commit_{parent1}$, visando obter as ações que foram feitas no ramo 1. Logo, as ações do ramo 1 são formalmente definidas como:

$$actions_{branch1} = diff(commit_{base}, commit_{parent1})$$

Por fim, é feito o *diff* entre $commit_{base}$ e $commit_{parent2}$, retornando as ações feitas no ramo 2. Portanto, as ações do ramo 2 são definidas formalmente como:

$$actions_{branch2} = diff(commit_{base}, commit_{parent2})$$

A partir dessas ações, é possível identificar as ações que representam retrabalho, trabalho desperdiçado e trabalho extra. Para o cálculo do retrabalho é feita a interseção entre as ações feitas nos ramos. O retrabalho é definido formalmente como:

$$actions_{rework} = actions_{branch1} \cap actions_{branch2}$$

A fim de determinar o trabalho desperdiçado e o trabalho extra é necessário primeiro identificar as ações que foram feitas em algum dos ramos. Portanto, definimos as ações dos ramos formalmente como:

$$actions_{branches} = actions_{branch1} + actions_{branch2}$$

Para calcular o trabalho desperdiçado, usamos o complemento relativo das ações de *merge* nas ações dos ramos, formalmente definido como:

$$actions_{wasted} = actions_{branches} \setminus actions_{merge}$$

Por fim, para determinar o trabalho extra usamos o complemento relativo das ações dos ramos nas ações de merge, formalmente definido como:

$$actions_{extra} = actions_{merge} \setminus actions_{branches}$$

Desta forma, é possível obter as métricas relativas a essas ações tanto de forma absoluta quanto de forma normalizada, conforme descrito a seguir:

$$rework = |actions_{rework}|$$
$$rework_{normalized} = rework \ / \ |actions_{branch1} \cup actions_{branch2}|$$
$$wasted = |actions_{wasted}|$$
$$wasted_{normalized} = wasted \ / \ |actions_{branches}|$$
$$extra = |actions_{extra}|$$
$$extra_{normalized} = extra \ / \ |actions_{merge}|$$

## 3. Implementação

Os conceitos discutidos na seção 2 foram implementados em uma ferramenta, que permite a coleta automática das métricas anteriormente apresentadas. A ferramenta foi desenvolvida em Python 3.6 e está disponível em https://github.com/gems-uff/merge-effort. O usuário pode utilizar a ferramenta de duas formas: via API ou através de um interpretador de linha de comando. Para ilustrar a utilização da ferramenta nessa seção, utilizamos o interpretador de linha de comando.

O usuário pode analisar tanto um repositório remoto, por exemplo, disponível no GitHub, quanto um repositório local, que já esteja em seu computador. Caso o usuário não utilize um repositório local, a ferramenta clonará o repositório temporariamente para fazer as análises. A ferramenta também disponibiliza a opção do usuário determinar uma lista de *commits* a serem analisados. Por default, caso o usuário não informe os *commits*, a ferramenta analisará todos os *commits* daquele repositório. Além disso, é possível informar se deseja obter as métricas absolutas ou normalizadas.

Criamos o exemplo da Figura 3 (a) para ilustrar o processo do reconhecimento dos esforços explicados na Seção 2. Esse exemplo faz uso de uma função fatorial intencionalmente simples, visando facilitar o entendimento da técnica proposta. A Figura 3 (b) apresenta o resultado dos *diffs* discutidos na Seção 2. Através do exemplo obtemos os seguintes esforços:

$$actions_{rework} = \{\text{"- if i<1:", "- fat(4)", "+ fat\_4= fat(4)", "+ print(fat\_4)"}\}$$

$$actions_{wasted} = \left\{\begin{array}{l}\text{"+ fat\_4= fat(4)", "+ fat\_4= fat(4)", "- if i< 1:",}\\ \text{"- fat(4)", "+ print(fat\_4)", "+ if i<= 1"}\end{array}\right\}$$

$$actions_{extra} = \{\text{"-def fat(i): ", "+ def fat\_iterativo(i):", "+ fat= fat\_iterativo(4)"}\}$$

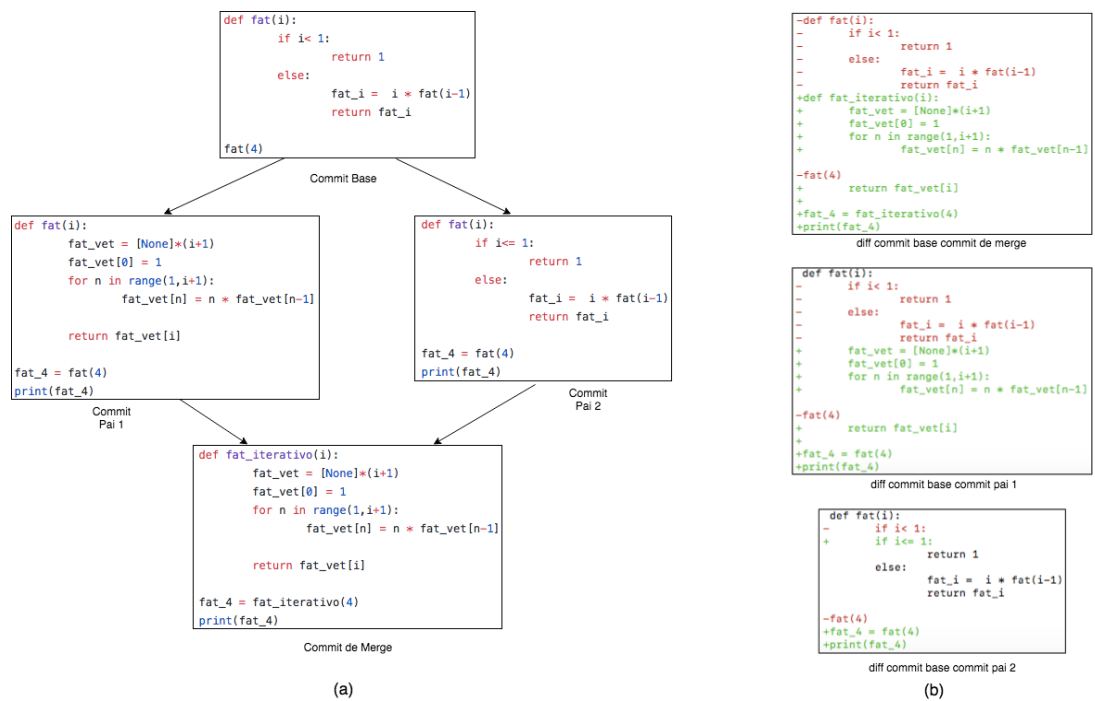

**Figura 3 Exemplo de merge e seus diffs**

Após todos os cálculos, a ferramenta retorna os valores de retrabalho, trabalho desperdiçado e trabalho extra. Para o exemplo, a saída da ferramenta é: 14 ações realizadas pelo ramo 1, 5 ações realizadas pelo ramo 2, 16 ações realizadas no *merge*, 4 ações de retrabalho, 6 ações de trabalho desperdiçado e 3 ações de trabalho extra.

## 4. Avaliação

Com o propósito de avaliar a utilidade da nossa ferramenta analisamos cinco repositórios *open-source* obtidos no GitHub. Utilizamos os seguintes critérios para a escolha dos projetos: ter pelo menos 1.000 *commits* e pelo menos 10 contribuidores. A Tabela 1 contém os cinco projetos selecionados, o número total de *commits*, o número total de contribuidores, a linguagem e a quantidade de *commits* de *merge*.

**Tabela 1 Projetos Selecionados**

| Projeto | Commits | Contribuidores | Linguagem | Commits de Merge |
|---|---|---|---|---|
| Sapos | 1.344 | 11 | Ruby | 153 |
| Voldemort | 4.262 | 61 | Java | 495 |
| DBAAS | 5.079 | 18 | Python | 677 |
| Jquery | 6.327 | 271 | JavaScript | 250 |
| Firefox-ios | 6.936 | 114 | Swift | 1.505 |

Através da análise dos resultados da nossa ferramenta para os projetos selecionados, construímos a Tabela 2, onde temos a porcentagem de *commits* que têm cada um dos esforços em relação à quantidade de *commits* de *merge* e a média de ações de cada esforço em relação à quantidade de *commits* de *merge*. Concluímos que o trabalho extra (*extra*) é o que tem menor ocorrência, acontecendo entre 0,47% a 12,42% dos casos. Vale notar que esses valores estão alinhados com o que é discutido na literatura [Menezes 2016], onde 13% dos *chunks* em conflitos foram resolvidos com código novo, ou seja, houve esforço adicional. Em contrapartida, o retrabalho (*rework*) e trabalho desperdiçado (*wasted*) ocorrem com mais frequência, entre 8,7% a 30,4% e 2,06% a 31,37%, respectivamente. É importante ressaltar que o projeto Firefox-ios é um *outlier* quando comparado aos outros. Se o excluíssemos da análise, obteríamos 16,54% a 30,4% de retrabalho e 11,96% a 31,37% de trabalho desperdiçado.

**Tabela 2 Resultados de retrabalho, trabalho desperdiçado e trabalho extra**

| Projeto | Commits de Merge | | | Média de Ações | | |
|---|---|---|---|---|---|---|
| | Rework | Wasted | Extra | Rework | Wasted | Extra |
| Sapos | 30,07% | 31,37% | 12,42% | 72,80 | 78,01 | 6,49 |
| Voldemort | 24,04% | 18,99% | 7,68% | 101,02 | 178,66 | 11,72 |
| DBAAS | 16,54% | 11,96% | 4,14% | 19,72 | 21,31 | 1,82 |
| Jquery | 30,40% | 21,20% | 8,00% | 22,89 | 23,21 | 0,42 |
| Firefox-ios | 8,70% | 2,06% | 0,47% | 3,20 | 3,60 | 0,25 |

Em relação a análise de média de ações de esforço, é possível notar que a quantidade média de ações de trabalho extra é significativamente menor que as demais. Além disso, observamos um padrão entre a quantidade média de ações de retrabalho e de trabalho desperdiçado, já que os valores são próximos. Por fim, um outro padrão observado é que a quantidade média de ações de retrabalho é sempre menor do que a de trabalho desperdiçado. Contudo, uma análise mais profunda e abrangente deve ser feita para obter evidências mais fortes sobre tais padrões e identificar as suas razões e implicações.

## 5. Trabalhos Relacionados

Prudêncio et al. [2012], com o intuito de reduzir a quantidade de conflitos, criaram a ferramenta Orion para avaliar qual é a política de controle de concorrência mais adequada para cada elemento de software. Para esse fim, a ferramenta calcula duas métricas: *merge effort* e *concurrency*. Apesar de utilizar o cálculo de esforço de merge,

o objetivo da ferramenta não é retornar as métricas de esforço para o usuário da forma que nossa ferramenta faz, mas sim propor uma política de controle de concorrência. Posteriormente, Santos et al. [2012] realizaram um estudo em que propõem dez métricas diferentes para a estimar o esforço de *merge*. Dentre as métricas a que obteve melhor resultado foi *Quantidade de Conflitos Físicos*. Além disso, Santos et al. [2012] apresentaram métricas que se baseiam na quantidade de linhas modificadas. Ambos os trabalhos fazem uso de conjuntos para o cálculo de esforço de merge, e não multisets como propomos nesse trabalho, levando a potenciais erros quando as mesmas ações acontecem em paralelo. Além disso, os estudos não abordam a métrica de retrabalho.

Em um outro trabalho, Mehdi et al. [2014] propuseram a análise de *merge* através da análise de duas métricas: *merge blocks* e *merge lines*. *Merge blocks* representa a quantidade de blocos modificados para resolver o conflito e *merge lines* representa a quantidade de linhas modificadas nos blocos. O objetivo da ferramenta criada por Mehdi et al. [2014] é utilizar as métricas para poder comparar o quão bom ou ruim pode ser um algoritmo de *merge*. Portanto, a ferramenta não tem o objetivo de retornar ao usuário as métricas de esforço, mas somente utilizá-las como um meio para avaliar os algoritmos de *merge*.

Em um estudo mais recente, Cavalcanti et al. [2017] analisou os esforços de *merge* semi-estruturado e *merge* não estruturado (convencional) através da quantidade de conflitos falsos positivos, o tempo necessário para solucionar o conflito de *merge* e o tempo necessário para pensar em como solucionar esse conflito. Na pesquisa, foi proposta uma ferramenta para análise de *merge* semi-estruturado com o intuito de reduzir o número de conflitos falso positivos e falso negativos. Novamente, a ferramenta sugerida não tem o propósito de retornar ao usuário as métricas de esforço de *merge*. Além disso, as métricas sugeridas não permitem a construção de uma ferramenta automática, já que não há como quantificar automaticamente com precisão o tempo que um desenvolvedor leva para pensar em como solucionar um conflito.

Além dos pontos já abordados, nenhum dos estudos tem a sua técnica de cálculo de esforço de *merge* detalhada o suficiente para permitir a sua reprodutibilidade.

## 6. Conclusão

Existem estudos que retrataram o esforço de merge, mas não foi encontrado um estudo cujo foco era a coleta de métricas de esforço e que relatasse com detalhes a sua técnica. Além disso, nenhum dos estudos relacionados propôs o uso de *multisets* para representar o conjunto de ações, o que é o mais indicado, tendo em vista que é possível a repetição de ações.

Esse estudo expôs, detalhadamente, uma técnica para extração de três métricas: retrabalho, trabalho desperdiçado e trabalho extra, além de apresentar uma implementação para essa técnica. Exibimos, também, uma análise feita com a nossa ferramenta onde pudemos concluir que o trabalho extra não acontece com muita frequência. Além disso, identificamos que o retrabalho e o trabalho desperdiçado acontecem entre 10% a 30% dos *commits* de merge. Até onde temos conhecimento, nenhum outro trabalho estudou a frequência de ocorrência de retrabalho ou trabalho desperdiçado. Também foram analisadas as quantidades médias de ações de cada tipo de esforço e concluiu-se que também há uma média maior de ações de trabalho desperdiçado e retrabalho.

Optamos por fazer a coleta das métricas no nível físico por ser o nível mais usado pelos controles de versão atuais, mas essa escolha nos leva a limitações. Dessa forma a ferramenta pode retornar falsos positivos, como por exemplo, o caso em que o usuário troca *tab* por espaços. Por isso, em um trabalho futuro, pretendemos adaptar a ferramenta para que possa fazer análise sintática, no nível da Árvore Sintática Abstrata.

Por fim, vemos como trabalho futuro um estudo mais profundo entre a possível correlação entre retrabalho e trabalho desperdiçado. Além disso, pretendemos estender a nossa análise para um universo maior de projetos. Por fim, estudaremos as possíveis razões para a ocorrência de retrabalho, trabalho desperdiçado e trabalho extra.

**Referências**